\documentclass[preprint]{aastex}

\shorttitle{Formation of Jovian and Saturnian Satellites}
\shortauthors{Sasaki, Stewart, and Ida}

\begin{document}

\title{Origin of the Different Architectures of the Jovian and Saturnian Satellite Systems}

\author{T. Sasaki}
\affil{Earth and Planetary Sciences, Tokyo Institute of Technology, 2-12-1 Ookayama, Meguro-ku, Tokyo 152-8551, Japan}
\email{takanori@geo.titech.ac.jp}

\author{G. R. Stewart}
\affil{Laboratory for Atmospheric and Space Physics, University of Colorado, Campus Box 392, Boulder, CO 80309-0392, USA}
\email{gstewart@lasp.colorado.edu}

\and

\author{S. Ida}
\affil{Earth and Planetary Sciences, Tokyo Institute of Technology, 2-12-1 Ookayama, Meguro-ku, Tokyo 152-8551, Japan}
\email{ida@geo.titech.ac.jp}

\begin{abstract}
The Jovian regular satellite system mainly consists of four Galilean satellites that have similar masses and are trapped in mutual mean motion resonances except for the outer satellite, Callisto.  On the other hand, the Saturnian regular satellite system has only one big icy body, Titan, and a population of much smaller icy moons.  We have investigated the origin of these major differences between the Jovian and Saturnian satellite systems by semi-analytically simulating the growth and orbital migration of proto-satellites in an accreting proto-satellite disk. We set up two different disk evolution/structure models that correspond to Jovian and Saturnian systems, by building upon previously developed models of an actively-supplied proto-satellite disk, the formation of gas giants, and observations of young stars. Our simulations extend previous models by including the (1) different termination timescales of gas infall onto the proto-satellite disk and (2) different evolution of a cavity in the disk, between the Jovian and Saturnian systems.
We have performed Monte Carlo simulations and
show that in the case of the Jovian systems,
four to five similar-mass satellites are likely to 
remain trapped in mean motion resonances.
This orbital configuration is formed by type I migration, 
temporal stopping of the migration near the disk inner edge, 
and quick truncation of gas infall caused by Jupiter opening a gap in the Solar nebula.  
The Saturnian systems tend to end up with one dominant body 
in the outer regions caused by the slower decay of gas infall 
associated with global depletion of the Solar nebula.  
The total mass and compositional zoning of the predicted Jovian and Saturnian 
satellite systems are consistent with the observed satellite systems.

\end{abstract}

\keywords{planets and satellites: formation --- planets and satellites: individual (Galilean satellites, Titan) --- planets: rings}

\section{INTRODUCTION}

The four Galilean satellites around Jupiter have similar masses and the inner three bodies are trapped in mutual mean-motion resonances.  
They exhibit a trend of decreasing mean bulk density with increasing orbital radius, and a diversity of axial moments of inertia (Table 1).  These properties are likely caused by a progressive increase in ice-to-rock ratio 
with an orbital radius and different states of differentiation.  
The inferred undifferentiated interior of the outermost satellite, Callisto, would require a relatively long accretion timescale, $5\times 10^5$ years or more (Stevenson et al. 1986; Schubert et al. 2004; Barr \& Canup 2008).  On the other hand, Saturn has only one big icy satellite, Titan, which is located relatively far from Saturn.  Recent Cassini data indicates that Titan also has an incompletely diﬀerentiated interior (Iess. et al. 2010) (Table 1).
The origin of the pronounced difference between the Jovian and Saturnian systems is an intriguing question. 

Two important models for circum-planetary proto-satellite disks have been proposed.
One is the ``solids enhanced minimum mass" (SEMM) model (Mosqueira \& Estrada 2003a, 2003b; Estrada et al. 2009) 
and the other is an actively-supplied gaseous accretion disk (CW) model (Canup \& Ward 2002, 2006, 2009).
The SEMM model postulates a massive quiescent disk with a peak surface density near $10^5$ g cm$^{-2}$ and a temperature profile determined by the luminosity of the gas giant planet. The Jovian/Saturnian satellites in this model form from solid materials supplied by ablation and capture of planetesimal fragments passing through the massive disk (Mosqueira et al. 2010).  The CW model postulates a low mass, viscously evolving disk with a peak surface density near 100 g cm$^{-2}$, that is continuously supplied by mass infall from the circum-stellar proto-planetary disk onto the circum-planetary proto-satellite disk.  The disk temperature profile in the CW model is dominated by viscous heating of the evolving disk and the luminosity of the giant planet plays a lesser role.  The ``satellitesimals" in this model are assumed to form immediately from dust grains that are supplied by gas infall (Canup \& Ward 2006).  Since the motions of the ``satellitesimals" are decoupled from gas accretion, they are retained while the disk gas is actively replenished.
As a result, a high dust-to-gas ratio is realized in the disk.
The satellites that have grown massive enough are disposed through type I migration into the planet
due to satellite-disk interaction (e.g., Ward 1986; Tanaka et al. 2002)
and new generations of satellites are repeatedly accreted from the supplied dust grains.
The finally surviving satellites are those formed at the very end of 
the host planet's accretion.
In the present paper, we will focus on the latter model. 

Canup \& Ward (2006) performed N-body simulations of accretion of satellites from small bodies
and their type I migration in decaying accretion disks. 
They showed that the equilibrium total mass of satellites resulting from a balance between disposal
by type I migration and repeated satellitesimal accretion
is universally $\sim 10^{-4} M_p$, where $M_p$ is the host planet mass.
The proto-satellite disk gas is eventually depleted 
due to the decline of infall gas due to global depletion of
the proto-planetary disk, then type I migration stops and
satellites with total mass $\sim 10^{-4} M_p$ survive.
The result is consistent with actual Jovian and Saturnian systems.
They suggested that the difference between Jovian and Saturnian systems
is caused by stochastic timing between the repeated formation/migration
and depletion of disk gas.

The disk evolution in their model is regulated by the mass infall rate 
from the proto-planetary disk onto the proto-satellite disk 
(section 2.1) and
the infall is responsible for the final stage gas giant planet formation.
Although Canup \& Ward (2006) assumed the same proto-satellite disk evolution
for both Jovian and Saturnian systems,
recent theories of gas giant formation and observational data 
of extrasolar planets raise the possibility 
that the disk evolution can be different between 
Jovian and Saturnian systems as follows.

Unless the growth of
gas giants is truncated or at least significantly slowed down 
at some critical mass, only a single gas giant should exist in each 
planetary system, because gas accretion onto a planet is a runaway process
(e.g., Bodenheimer \& Pollack 1986; Pollack et al. 1996; Ikoma et al. 2000).
Cores for gas giants do not generally 
form at the same time, and the rate of gas 
mass accretion in the proto-planetary disk 
that can supply gas to giants
is observationally $\sim$ O$(10^{-5}) M_J$ year$^{-1}$
(where $M_J$ is a Jupiter mass). 
However, the observed data of extrasolar planets show 
that there are many systems 
with multiple gas giants.
Our Solar system also has two gas giants.
Furthermore, the mass distribution of extrasolar gas giants 
is centered at $M_{p,ave} \sim \mbox{a few} M_{\rm J}$ 
with an upper cut-off of $M_{p,max} \sim 10 M_{\rm J}$.
The observationally inferred proto-planetary disk masses
have a mean value of $\sim 10 M_J$ 
and a maximum value of $\sim 100 M_J$ (e.g., Beckwith \& Sargent 1996).
These mean and maximum values are
one order greater than $M_{p,ave}$ and $M_{p,max}$, respectively.
These facts 
suggest that the giants did not accrete 
most of disk gas, so they
are consistent with existence of the rapid truncation 
(e.g., Lin \& Papaloizou 1985) or 
severe decline 
(e.g., Lubow \& D'Angelo 2006; Tanigawa \& Ikoma 2007)
of gas infall onto the planets due to gap opening
in the proto-planetary disks.

Since the proposed gap opening conditions show that 
the critical mass for gap opening is larger in outer regions
(section 2.2),
outer gas giant planets, in principle, tend to be larger than inner ones.
However, the outer one can be smaller, if the disk gas is depleted before the outer planet can complete its formation.
Since Saturn is three times less massive than Jupiter, it is
most likely that Jupiter opened up a gap to halt its growth while
Saturn did not and its growth was terminated by
global depletion of the proto-planetary disk (the Solar nebula).
Although reduced gas infall can still continue 
after the gap opening (Lubow \& D'Angelo 2006; Tanigawa \& Ikoma 2007),
the severe reduction in infall rate makes 
the Jovian disk evolution significantly different from
the Saturnian one (see section 2). 

Here, we explore a possible path to produce the pronounced different 
architectures of the Jovian and Saturnian satellite systems, by introducing
the different evolution of proto-satellite disks due to 
the gap opening to the actively-supplied gaseous accretion proto-satellite 
disk model (Canup \& Ward 2002, 2006).
The purpose of the present paper is to demonstrate how this diversity in gas giant formation can profoundly affect the regular satellite formation process.

Since we survey a wide range of parameters for initial and boundary conditions, 
we adopt a semi-analytical model to simulate accretion 
and migration of satellites 
from satellitesimals that has been developed by Ida \& Lin (2004, 2008)
for modeling sequential planet formation.
While the semi-analytical calculations inevitably introduce
approximations, N-body simulations have to employ 
unrealistic initial and/or boundary conditions 
because of the computational limitations (e.g., Kokubo \& Ida (1998) started 
from planetesimals of $\sim (10^{-4}-10^{-3}) M_{\oplus}$, 
or Canup \& Ward (2006) replaced infalling dust grains by embryos with
the isolation mass of an accreting satellite).
We show in section 4.1 that our calculation produces results 
consistent with
Canup \& Ward (2006)'s N-body simulation, although
inconsistency may exist in detailed features
(we need more careful comparison to reconcile the inconsistency).

In section 2, we explain our disk evolution models for Jovian and
Saturnian systems in details, highlighting their differences.
In section 3, semi-analytical treatments to simulate 
accretion of satellites from satellitesimals, orbital migration,
and resonant trapping are described.
The simulation results are presented in section 4.
We successfully explain the pronounced difference between Jovian
and Saturnian systems in the framework of our model.
We also discuss implications for the formation of Saturn's rings.  
Section 5 is devoted to a summary of the major results.

\section{SATELLITE FORMATION CIRCUMSTANCES}

We modify the actively-supplied disk model developed by 
Canup \& Ward (2002, 2006) (see section 2.1),
by introducing the effects of the gap formation around 
a host planet's orbit.
If a gap is opened, the
proto-satellite disk rapidly dissipates (section 2.2), 
so that the mass distribution and orbital configuration of satellites
are ``frozen" in the state of a relatively massive and hot disk.
The hot disk condition allows retention of rocky satellites corresponding to
Io and Europa.
The massive disk condition could also result in formation of an inner cavity of
the disk (Takata \& Stevenson 1996; also see section 2.4)
that plays a crucial role in formation of resonantly trapped 
multiple satellites like the Galilean satellites.

Observations of Classical T-Tauri Stars (CTTSs) with
relatively strong mass accretion and Weak line T-Tauri Stars (WTTSs) 
with relatively weak mass accretion suggest that
magnetic coupling between disks and
host stars becomes weak from CTTSs to WTTSs.
 It may imply that the inner cavity would vanish 
as disk surface density and disk accretion rate decay,
corresponding to evolution from CTTS stage to WTTS one.
We adopt the same picture for evolution 
of proto-satellite disks (section 2.4).

The Saturnian disk would also have experienced the massive 
disk stage with an inner cavity to have resonantly trapped multiple 
satellites.
However, if Saturn did not open up a gap in the protoplanetary disk, then
as the cavity vanishes, the massive satellites are released 
to migrate onto the host planet.
The finally surviving satellites are therefore generally formed in depleted and cold disk conditions, 
in which only ice-rich bodies exist (section 2.3).

\subsection{Actively-supplied disk model} 

First we briefly summarize descriptions for evolution of
the actively-supplied accretion disk model with relatively small mass 
$\sim 10^{-4} M_p (p = Jupiter, Saturn)$ developed 
by Canup \& Ward (2002, 2006), which we follow in the present paper.
More detailed descriptions are given in Appendix A.

Gas infall from the proto-planetary disk 
onto the proto-satellite 
disk is limited to the regions inside a critical radius $r_c$.
Canup \& Ward (2006) assumed that the infall flux
inside $r_c$ is independent of orbit radius and $r_c$ has a fixed value of 30$R_p$, where $R_p$ is the physical radius of the host planet. (Realistically, the infall mass flux can have a radial dependence and should evolve with time, but current hydrodynamic simulations of gas giant planet formation lack sufficient resolution and long enough integration times to constrain the radial and time dependence of the infall.)
That is, the infall flux per unit area has a constant value of 
$(F_p/\pi r_c^2)$ at $r < r_c$ and vanishes at $r > r_c$, where
$r$ is the orbital radius around the host planet and
$F_p$ is the total mass infall rate onto the disk.
In the steady accretion state of the proto-planetary disk, 
the infall rate is
defined with a parameter $\tau_G$ as $F_{p,0} = M_p/\tau_G$.
Considering non-uniform inflow distributions would be an important area of future investigation.

Neglecting a diffused-out extended faint disk 
to make clear dynamical process of
accretion and migration of proto-satellites, surface density of disk gas is
given by (Appendix A)
\begin{equation}
\Sigma_g \simeq 
\left\{
\begin{array}{ll}
{\displaystyle
0.55\frac{F_p}{3\pi\nu} 
\simeq 100f_g\left(\frac{M_p}{M_J}\right)\left(\frac{r}{20R_J}\right)^{-3/4}
\mbox{g cm$^{-2}$}} & [r < r_c], \\
0 & [r > r_c], 
\end{array}
\right.
\label{eq:sigma_g}
\end{equation}
where 
\begin{equation}
f_g \equiv \left(\frac{\alpha}{5\times 10^{-3}}\right)^{-1}\left(\frac{\tau_G}{5\times 10^6\mbox{yrs}}\right)^{-3/4},
\end{equation}
where we used the turbulent viscosity of the proto-satellite disk with 
$\nu = \alpha H_{\rm SD}^2 \Omega_{\rm K}$ (Shakura \& Sunyaev 1973),
$\Omega_{\rm K} = \sqrt{GM_p/r^3}$, 
and temperature distribution to calculate 
the disk scale height ($H_{\rm SD}$)
is given by eq.~(3).
Since $\nu \propto T_d \propto F_p^{1/4}$ (eq.~[3]),
$\Sigma_g \propto F_p/\nu \propto F_p^{3/4} \propto \tau_G^{-3/4}$.
After a quasi-steady state of accretion 
and destruction of satellites
(see section 2.2 and 2.3) is established, 
we start the exponential decay of infall as
$F_p = F_{p,0} \exp(-t/\tau_{\rm dep})$ with 
$\tau_{\rm dep} = (3$--$5)\times 10^6$ years, 
corresponding to global depletion of the proto-planetary disk.

We use the disk temperature in the steady state that is
derived by a balance between
viscous heating and black-body radiation (Appendix A),
\begin{equation}
T_{d,0} \simeq 160\left(\frac{M_p}{M_J}\right)^{1/2}
\left(\frac{\tau_G}{5\times 10^6 \mbox{yrs}}\right)^{-1/4}
\left(\frac{r}{20R_J}\right)^{-3/4}\mbox{K}.
\label{eq:temp_d}
\end{equation}
The disk is assumed to be vertically optically thin
and isothermal (Appendix A).
After the exponential decay of $F_p$ is imposed,
the disk temperature is decreased as
\begin{equation}
T_d = T_{d,0} \exp \left(-\frac{t}{4 \tau_{\rm dep}}\right).
\end{equation}

The disk structure (gas surface density and temperature) 
is therefore regulated by $F_p$ ($= M_p/\tau_G$), in the
actively-supplied disk model.
We discuss how $F_p$ is likely to evolve with time for Jovian and
Saturnian disks in the following.
We will also point out that the disk inner boundary condition 
may also be regulated by $F_p$ (section 2.4).

\subsection{Jovian System with Gap Opening}
 
Except for the late stages, growth of a gas giant is
a runaway process of accreting gas from the disk.
The mass increase is regulated by Kelvin-Helmholtz contraction
with a timescale (see e.g., Ida \& Lin 2004),
\begin{equation}
\tau_{\rm KH} = \frac{M_p}{\dot{M}_p} 
  \simeq 10^6 \left(\frac{M_p}{10 M_{\oplus}}\right)^{-3}{\rm years},
\end{equation}
which can have different numerical factors
depending on opacity, but the runaway nature is robust.
Then, $F_p \sim 3 \times 10^{-6} (M/0.1M_J)^4 M_J/{\rm yr} \propto t^{4/3}$.
After the planet accretes all the gas in feeding zone with width
$\sim 2 r_{\rm H}$, where Hill radius $r_{\rm H} = (M_p/3M_*)^{1/3} r$,
at $M_p \sim 0.3 M_J$ in the case of the Solar nebula, 
the growth would be regulated by supply of global disk accretion,
$\dot{M}_{\rm disk}$, which is $\sim 10^{-5} M_J$ year $^{-1}$
and decays on timescales of $10^6$--$10^7$years (see Fig.~1).  

As we will discuss in section 2.4, satellite formation 
would attain a quasi-steady state between accretion of satellites
from infalled materials and their migration onto the host planet.
Since we are concerned with the architecture of satellite systems
formed by the last survivors, we start calculations 
from relatively late stage with $\tau_G = 2\times 10^6$ years 
($F_p = 5 \times 10^{-7} M_J$ year $^{-1}$) for Jupiter. 
Although the choice of initial value of $\tau_G$ contains 
uncertainty, the critical satellite mass to start migration
and disk temperature depend on $\tau_G$ only weakly:
$M_{\rm mig} \propto \tau_G^{-3/16}$ (eq.~[15])
and $T_d \propto \tau_G^{-1/4}$ (eq.~[3]).
So, the mass distribution and compositions of final
satellites do not sensitively depend on the initial 
value of $\tau_G$. 
Furthermore, since Canup \& Ward (2006)'s N-body simulations adopted
similar values, we can directly compare 
our results with theirs.

As discussed in section 1, we assume that Jupiter 
opened up a gap in the Solar nebula.
Then, $F_p$ quickly decays (eq.~[9]).
It is proposed that a gap is opened if both viscous and thermal conditions
are satisfied (Lin \& Papaloizou 1985; Crida et al. 2006).
The former condition
is derived by comparison between the viscous diffusion of the gas
and gravitational scattering by the planet and the critical planetary 
mass is given by (Lin \& Papaloizou 1985; Ida \& Lin 2004)
\begin{equation}
M_{\rm g,vis} \simeq \frac{40 \nu}{r_p \Omega_{\rm K}}
M_* \simeq 40 \alpha \left(\frac{H_{\rm PD}}{r_p}\right)^2 M_*
\simeq 30 \left(\frac{\alpha}{10^{-3}}\right)
\left(\frac{r_p}{1{\rm AU}}\right)^{1/2}
\left(\frac{M_*}{M_{\odot}}\right)  M_{\oplus},
\label{eq:m_gas_vis}
\end{equation}
where $r_p$ is heliocentric orbital radius of the planet,
$\Omega_{\rm K}$ is Keplerian frequency at $r_p$, and
$M_*$ is the host star mass, 
For the scale height $H_{\rm PD}$ of the proto-planetary disk, we used 
the values of an optically thin disk (Hayashi 1981),  
\begin{equation}
H_{\rm PD} = 0.05 \left( \frac{r_p}{1{\rm AU}} \right)^{1/4} r_p.
\label{eq:h_a}
\end{equation}
The thermal condition is determined by balance between the gravitational
scattering and pressure gradient,
$r_{\rm H} \sim 1.5 H_{\rm PD}$ (Lin \& Papaloizou 1985; Ida \& Lin 2004).
The critical mass is given by
\begin{equation}
M_{\rm g,th}
\simeq 400 \left(\frac{r_p}{1{\rm AU}}\right)^{3/4}
\left(\frac{M_*}{M_{\odot}}\right) M_{\oplus}.
\label{eq:m_gas_therm}
\end{equation}
Note that $M_{\rm g,th}$ is generally larger than $M_{\rm g,vis}$,
because typical values of the $\alpha$ parameter are $\sim 10^{-3}$--$10^{-2}$ 
for turbulence induced by magneto-rotational instability 
(e.g., Sano et al. 2004).  

Since $M_{\rm g,th}$ is comparable to Jupiter mass, 
it has been proposed that Jupiter's final mass is determined by 
the gap opening (Lin \& Papaloizou 1985; Ida \& Lin 2004).
Recent fluid dynamical simulations suggest that the gap is not clear enough 
to halt gas accretion for $M_p \la 5M_J$ (e.g., Lubow et al. 1999; D'Angelo et al. 2003).
However, the gas flow across the gap is not necessarily
accreted by the planet (Dobbs-Dixon et al. 2007) and
even if all the passing flow is accreted by the planet, 
it is only $10\%$ of exterior disk mass accretion flow 
($\dot{M}_{\rm disk}$) or less for $M_p \sim M_J$
and quickly vanishes with increasing $M_p$
(Lubow \& D'Angelo 2006; Tanigawa \& Ikoma 2007).
We will show below that two orders of magnitude reduction
in the mass flux onto the proto-satellite disk is
enough to discriminate proto-satellite disk evolution with
the gap from that without the gap.

The truncation of gas infall or its severe decline due to the gap opening
depletes the circum-planetary proto-satellite disk 
on its own viscous diffusion timescale $\tau_{\rm diff} \sim R_{\rm SD}^2/\nu$ 
where $R_{\rm SD}$ is the disk size and $\nu$ is disk viscosity.
With the alpha model and $R_{\rm SD} \sim 20 R_{\rm p}$ (where
$R_{\rm p}$ is a physical radius of the planet), 
\begin{equation}
\begin{array}{ll}
\tau_{\rm diff} &
{\displaystyle
\sim \frac{R_{\rm SD}^2}{\alpha H_{\rm SD}^2 \Omega_{\rm K}} 
\sim \left( \frac{R_{\rm SD}}{H_{\rm SD}} \right)^2
     \frac{1}{\alpha} \frac{T_{\rm K}(R_{\rm SD})}{2 \pi} }\\
         &
{\displaystyle
\sim 10^3 \left(\frac{\alpha}{10^{-3}}\right)^{-1} {\rm years}},
\label{eq:t_diff}
\end{array}
\end{equation}
where $H_{\rm SD}$ is the scale height of the proto-satellite disk
and $T_{\rm K}$ is an orbital period around the planet given by
\begin{equation}
T_{\rm K}(r) = 0.032 \left(\frac{r}{20 R_p}\right)^{3/2} {\rm years}.
\end{equation}
The gap opening itself would occur on the viscous diffusion timescale for 
the scale height ($\sim H_{\rm PD}$) in the proto-planetary disk,
which is
\begin{equation}
\tau_{\rm gap} \sim \frac{H_{\rm PD}^2}{\nu} \sim 
\left( \frac{H_{\rm PD}}{R_{\rm PD}} \right)^2
\frac{R_{\rm PD}^2}{\nu} \sim (10^{-3}-10^{-4}) \times (1-10) {\rm Myrs}, 
\end{equation}
where $R_{\rm PD}$ is size of the proto-planetary disk, 
which would be much larger than 5AU, and $t_{\rm dep} \sim R_{\rm PD}^2/\nu$ 
is global disk evolution timescale.  
Both $\tau_{\rm diff}$ and $\tau_{\rm gap}$ are much shorter than 
formation and orbital evolution timescales of satellites (see section 3.1).  
This separation of time scales allows us to assume that the proto-satellite disk around Jupiter was 
abruptly depleted when the gap opened, such that the growth and orbital migration of satellites were ``frozen" at that time.

Even if we consider the imperfect truncation of infall,
this feature is not changed
as long as the infall rate is reduced by a factor larger than 100.
The equilibrium gas surface density of the proto-satellite disk ($\Sigma_g$)
is determined by infall flux ($F_p$) as $\Sigma_g \propto F_p^{3/4}$ 
(section 3.1).
If the infall rate is decreased by two-orders of magnitude, 
$\Sigma_{g}$ is decreased by a factor of 30.  
According to the decrease in $\Sigma_{g}$ and the associated decrease in
surface density of satellitesimals, accretion and migration timescales 
of proto-satellites become comparable to or longer than the global depletion 
timescale of the proto-planetary disk ($\sim 1$--10 Myrs; Meyer et al. 2007).  
Therefore, although the residual infall after the gap opening is considered, 
the ``frozen feature" is mostly preserved.
We also carried out runs with the residual continuous infall 
with a reduction factor of 100
and confirmed the above argument.
We will show that the introduction of the imperfect
truncation produces results that may be rather consistent with formation of
a Callisto analogue (section 4.3).

\subsection{Saturnian System without Gap Opening}

Both thermal (eq.~8) and viscous
(eq.~6) conditions show that the critical mass 
for gap opening is larger in outer regions of the solar system.
So, the final mass of Saturn should be larger than Jupiter's mass,
unless significant disk mass has already been accreted by the Sun and 
Jupiter such that the total residual disk mass is reduced to the level of Saturnian mass
when the Saturn's core is formed.
Since the core accretion from planetesimals is generally slower at larger orbital radius in the proto-planetary disk (e.g., Ida \& Lin 2004), it is reasonable that Saturn's core accreted after Jupiter has fully formed (Fig.~1).
Since it is most likely that Jupiter has not undergone significant 
type II migration, 
Jupiter may have formed in a partially depleted proto-planetary disk as well. 
Therefore, it is not unreasonable to assume that Saturn formed in a significantly depleted disk and did not open up a gap in the disk (Fig.~1).  
Actually, planets in further outer regions, Uranus and Neptune, 
have not undergone runaway gas accretion although their cores have large enough mass for runaway gas accretion.  This timing between core accretion and disk depletion 
would also produce diversity of extrasolar planets (e.g., Ida \& Lin 2008).

On the assumption that Saturn did not open up a gap, the Saturnian proto-satellite disk 
should have been gradually depleted on the global depletion timescales of the proto-planetary disk, 
$t_{\rm dep} \sim 1$--10 Myrs (Fig.~1).
The many orders of magnitude difference in depletion timescales
of the proto-satellite disk between Jovian ($\tau_{\rm diff} \sim 10^3$ years)
and Saturnian ($\tau_{\rm dep} \sim 10^6$--$10^7$ years) systems
would significantly affect the final configuration of satellite systems, 
because a typical type I migration timescale of proto-satellites 
($\tau_{\rm mig} \sim 10^5$ years; Tanaka et al. 2002) 
are in between the two timescales.   
Jovian satellites may retain their orbital configuration in a phase of the proto-satellite disk with relatively high mass that was
frozen at the time of abrupt disk depletion, 
while Saturnian satellites must be survivors against type I migration 
in the final less massive disk (Fig.~1).
The migration timescale depends on the magnitude
of infall flux ($F_p$) through $\Sigma_g$ and $\nu$, which remain uncertain.
However, since the migration timescale does not 
strongly depend on the flux ($\tau_{\rm mig} \propto F_p^{-1/2}$),
the relation of $\tau_{\rm diff} < \tau_{\rm mig} < \tau_{\rm dep}$
is not be changed by the uncertainty in $F_p$
even if the uncertainty is 2--3 orders of magnitude. 
We start calculations
from the last stage with $\tau_G = 5 \times 10^6$ years 
($F_p = 7 \times 10^{-8} M_J$ year $^{-1}$) for Saturn,
which is the same as the calculations by Canup \& Ward (2002, 2006).

\subsection{Difference in a Disk Inner Cavity}

Takata \& Stevenson (1996) studied {\rm magnetic} 
coupling between the proto-satellite disk 
and proto-Jupiter,
in order to address why Jupiter's spin rotation is substantially slower than
the break-up spin rate.
If Jupiter accretes disk gas rotating at Keplerian velocity
near its surface, Jupiter's spin rate should be close to
the break-up one.
If the coupling is strong enough, angular momentum is transferred from the
Jupiter's spin to the disk through the magnetic field.
In this case, it is expected that the disk is truncated at the corotation radius 
where the disk gas corotates with the planetary spin.
The critical magnetic field for the magnetic coupling 
is a few hundred Gauss for mass accretion rate onto a planet of $\sim 10^{-6} M_J$ year$^{-1}$,
if we use a standard formula of magnetospheric radius (K\"onigl 1991).  
Such a strong magnetic field for proto-Jupiter 
could be consistent with a scaling argument (Stevenson 1974).  
Actually, Takata \& Stevenson (1996) found through numerical calculation that
the proto-Jovian disk may have had an inner cavity created by the magnetic coupling.  

Although evolution of the inner cavity in the proto-satellite disk is not 
theoretically clear,
the observations for proto-planetary disks around T Tauri stars suggest
that the inner cavity exists in early active stage and 
vanishes as the disk accretion onto a central body
becomes weak.
The observed spin period distribution of young stars is
bimodal and CTTSs tend to rotate slower than WTTSs (Hartman 2002; Herbst \& Mundt 2005).  
A possible idea to account for the bimodality is as follows.
The relatively high disk accretion onto CTTSs may maintain
the magnetic field strong enough for the magnetic coupling
to transfer spin angular momentum to the disk and create a cavity.
On the other hand, with the lower disk accretion rate
in the following WTTS stage,
the coupling and hence the cavity vanishes and 
the host star accretes high angular momentum gas from the disk,
becoming a rapid rotator (Herbst \& Mundt 2005).  
Recent Spitzer observations also support this idea
(Rebull et al. 2006; Cieza \& Baliber 2007).
 
Based on the analogy of the evolution from CTTSs to WTTSs,
we here assume simulation conditions of 
(I) a truncated boundary with an inner cavity and
abrupt termination of infall for the Jovian system 
and (II) a non-truncated boundary without a cavity 
and gradual depletion of infall for the Saturnian system.
The detailed settings are described in section 3.1.

The assumption may include large uncertainty 
and needs to be studied in greater detail.
However, the purpose of the present paper is 
to demonstrate the important link
between gas giant planet formation and satellite formation through 
evolution of the proto-satellite disk due to changes in infall rate 
and magnetic coupling.

\section{METHODS FOR SIMULATIONS FOR SATELLITE FORMATION}

Given a disk evolution model,
we apply the population synthesis planet formation model (Ida \& Lin 2004, 2008) 
to simulate growth of proto-satellites through accretion of small bodies (``satellitesimals") and 
their inward orbital migration caused by tidal interactions with the gas disk. 

\subsection{Evolution of infall} 

In set II for Saturn, a quasi-steady state in which accretion of satellites
is balanced with their removal to the host planet via type I migration
is established.
After the quasi-steady state is established, we start
the exponential decay of infall as
$F_p = F_{p,0} \exp(-t/\tau_{\rm dep})$ with 
$\tau_{\rm dep} = (3$--$5)\times 10^6$ years.
We integrate the systems until
disk gas surface density decreases so much that
satellites no longer undergo orbital migration. 
This setting is the same as that adopted by Canup \& Ward (2002, 2006).  

In set I for Jupiter, we set the disk inner edge
at a corotation radius of Jupiter ($\sim 2.25R_p$).
In this case, satellites are piled up outside
the disk edge.
As we will argue in section 3.3, 
if the amount of trapped satellites exceeds
a critical value, the innermost one is released.
Thus, a quasi-steady state different from set II is
established also in set I.
After the establishment of the quasi-steady state,
we start the exponential decay of infall 
in the same way as in set II.
However, we stop integration randomly at 
(0.5--$0.8) \tau_{\rm dep}$ after the exponential 
decay starts, which reflects the very rapid depletion 
of the disk due to the gap opening.  
We also did calculations (set I') with
100 times reduced infall flux after the gap opening,
instead of the complete termination, which corresponds
to imperfect gap opening.

While the evolution of the gas surface density of the proto-satellite disk is 
analytically given by eq.~(1),
the surface density of satellitesimals is consistently calculated 
with removal due to accretion by proto-satellites and 
supply from solid components in the infalling gas. 
We scale the solid surface density with a scaling factor $f_d$ as
\begin{equation}
\Sigma_d = \eta_{\rm ice}f_d\left(\frac{M_p}{M_J}\right)\left(\frac{r}{20R_J}\right)^{-3/4} \mbox{g cm$^{-2}$},
\end{equation}
where $\eta_{\rm ice}$ is an enhancement factor due to condensation of icy grains at $r > r_{\rm ice}$ where disk temperature is lower than 160K; we adopt $\eta_{\rm ice}=3$ for $r>r_{\rm ice}$ and $\eta_{\rm ice}=1$ otherwise.  
In the infall flux, the mass fraction of solid components is given by 
$\eta_{\rm ice}/f$, where $f$ is the mass ratio of gas relative to 
silicate dust.
For a fiducial value, we adopt $f=100$.  
If $f_d=f_g$, the gas to dust ratio of the disk is the same as that of infall.
But, we found that $f_d$ is usually 10-100 times higher than $f_g$ in outer regions
in a steady state (section 4.1).
The method used to calculate the evolution of $\Sigma_d$ is described in Appendix A.

\subsection{Accretion and Migration of Proto-Satellites} 

We randomly select initial locations of 10-20 satellite seeds 
with $M = 10^{20}$ g ($\sim 10^{-11}M_J$, $\sim 10^{-10}M_S$) from a log-uniform distribution in the regions inside $r_c = 30R_p$ and calculate
their growth, migration, resonant trapping, and re-generation,
simultaneously calculating the evolution of $\Sigma_d$,
until disk gas is depleted.  
The log-uniform distribution for initial locations 
corresponds to the spacing between 
the seeds being proportional to orbital radius, 
which is the simplest and natural choice (Ida \& Lin 2004).  
Although the choice of the initial locations is arbitrary, 
it would not change the overall results thanks to 
the effects of type-I migration.  
The numbers of the initial satellite seeds ($N = 10$--$20$)
do not affect the mass and orbital distributions of final
satellites as well, 
as long as the initial mass is small enough,
because $\tau_{\rm acc}$ is shorter for smaller $M$ 
(eq.~13) and only a few bodies 
undergo runaway growth from the seeds.

The accretion rate of a proto-satellite from satellitesimals is given by 
$\dot{M} = M/\tau_{\rm acc}$ and the
accretion timescale is (Appendix B)
\begin{eqnarray}
\tau_{\rm acc} & \simeq & 
0.5\left(\frac{r}{R_p}\right) \left(\frac{M_p}{\Sigma_d r^2}\right)
\left(\frac{M}{M_p}\right)^{1/3} T_K \nonumber \\
 &\simeq& 10^6 f_d^{-1}\eta_{\rm ice}^{-1}
\left(\frac{M}{10^{-4}M_p}\right)^{1/3}\left(\frac{M_p}{M_J}\right)^{-5/6}
\left(\frac{r}{20R_J}\right)^{5/4} \mbox{years},
\label{eq:tau_acc}
\end{eqnarray}
where $M$ is the satellite mass.
The type I migration rate is $\dot{r} = r/\tau_{\rm mig}$ 
and the migration timescale is (Tanaka et al. 2002)
\begin{eqnarray}
\tau_{\rm mig} & \simeq & 0.3 \left(\frac{c_s}{r\Omega_K}\right)^2\frac{M_p}{M}\frac{M_p}{r^2\Sigma_g}\Omega_K^{-1} \nonumber \\
 &\simeq& 10^5 f_g^{-1}\left(\frac{M}{10^{-4}M_p}\right)^{-1}\left(\frac{M_p}{M_J}\right)^{-1}\left(\frac{r}{20R_J}\right)^{1/2}\left(\frac{\tau_G}{5\times 10^6\mbox{yrs}}\right)^{-1/4} \mbox{years} \nonumber \\
 &\simeq& 10^{5} \left(\frac{\alpha}{5 \times 10^{-3}}\right)\left(\frac{M}{10^{-4}M_p}\right)^{-1}\left(\frac{M_p}{M_J}\right)^{-1}\left(\frac{r}{20R_J}\right)^{1/2}\left(\frac{\tau_G}{5\times 10^6\mbox{yrs}}\right)^{1/2} \mbox{years}.
\label{eq:t_mig}
\end{eqnarray}
We do not include reversed torque due to a cavity (Masset et al. 2006)
nor entropy gradient (Paardekooper et al. 2009) for simplicity.
If a proto-satellite migrates to the host planet or collides with another proto-satellite,
a next-generation seed with mass $M = 10^{20}$ g
is generated in the regions out of which preceding runaway bodies have 
migrated leaving many satellitesimals. 

When $\tau_{\rm acc}$ becomes larger than $\tau_{\rm mig}$,
migration actually starts.
From eqs.~(13) and (14), the
critical satellite mass for the migration is given by
\begin{equation}
M_{\rm mig} \simeq 
2 \times 10^{-5} \eta_{\rm ice}^{-3/4}
\left(\frac{M_p}{M_J}\right)^{-1/8}
\left(\frac{r}{20R_J}\right)^{-9/16} 
\left(\frac{\tau_G}{5\times 10^6\mbox{yrs}}\right)^{-3/16} M_*.
\label{eq:m_mig}
\end{equation}

\subsection{Trapping by Resonances and Disk Edge}

When the orbits of proto-satellites approach each other 
due to their differential type I migration speed, 
angular momentum is transferred between them through mutual 
gravitational perturbations and they are often trapped 
in mean motion resonances.  
As explained in Appendix C, migration in proto-satellite disks
is slow enough that
resonant trapping occurs with very high probability.
We assume that the two approaching proto-satellites are 
trapped at a resonance near an orbital separation of $5r_{\rm H}$.  

For the Saturn case, we will show that large proto-satellites generally
form in the outer regions and they sweep up smaller
proto-satellites in the inner regions with the resonant 
trapping, because larger bodies migrate faster.

For the Jupiter case, migration is halted
at the disk inner edge and subsequently migrating satellites 
are trapped in resonances one after another.
Individual satellites, except the innermost one,
should keep 
losing angular momentum through type-I migration 
even after the trapping.
However, Ogihara \& Ida (2009) used N-body simulations that included the damping due to disk-planet interactions to produce resonantly trapped satellites that are lined up outside the inner disk edge with the innermost one pinned up at the edge even in the case without the effect of reversed type I migration torque near the edge (Masset et al. 2006).
Ogihara et al. (2010) has investigated this halting
mechanism in detail and found that asymmetric eccentricity
damping for the body in an eccentric orbit straddling 
the edge transfers angular momentum from the 
disk to the body.
Since the timescale of the eccentricity damping is
much shorter than type I migration timescale
by a factor of $\sim (c_s/v_{\rm K})^2 \sim$ O$(10^{-3})$
(Tanaka et al. 2002; Tanaka \& Ward 2004), this angular momentum supply is
great enough to compensate for the loss due to type I migration
torques on all the trapped bodies (for details, 
see Ogihara et al. (2010)).
For the angular momentum supply mechanism to
work, eccentricity of the innermost body must
be continuously excited.
In the case of the resonantly trapped bodies, the eccentricity of the innermost body is continually excited by resonant interactions with the other bodies in the resonance.
The maximum number of trapped satellites may be several,
if their formula is applied.

However, there is another limitation for the trapping,
which is not considered in Ogihara et al. (2010).
When the total mass of the trapped satellites exceeds 
the total disk mass, 
angular momentum loss due to type I migration 
is absorbed by outward diffusion 
of the disk to outer regions
in which the specific orbital angular momentum is higher.
As a counter-reaction,
the inner edge is no longer able to halt the satellites' inward migration.
The innermost satellite is therefore released to the planetary surface.
Thus, Jupiter systems also attain a quasi-steady state 
in which the total mass in trapped satellites is kept
almost constant with small amplitude oscillations.
We here adopt the latter condition rather than
the condition derived by Ogihara et al. (2010).
(The former condition is much more important for studying
formation of close-in super-Earths in extrasolar
planetary systems; Ida \& Lin (2010).)

We summarize the simulation settings for Jovian (set I)
and Saturnian (set II) systems in Table II.

\subsection{Comparison with N-Body Simulations}

To check the validity of our semi-analytical model, 
we carried out the calculations with the same disk conditions 
as those in Fig.~2 of Canup \& Ward's N-body simulations (Canup \& Ward 2006).  
Here $F_p$ is constant without the exponential decay.
The non-cavity condition and $f=100$ are adopted for a Jupiter mass planet.  
The total mass in satellites, $M_T$, in our simulation
is plotted as a function of time 
for $\alpha = 10^{-4}$ (solid line), $\alpha = 5\times 10^{-3}$ (dashed line), 
and $\alpha = 5\times 10^{-2}$ (dotted line) in Fig.~2.  
After an initial buildup, an equilibrium state for $M_T$ is attained in which disposal of proto-satellites to the host planet due to type I migration is balanced with the supply of satellitesimals from the infall.  
The dotted horizontal lines are Canup \& Ward (2006)'s analytical estimates for
individual values of $\alpha$ with which the asymptotic values of $M_T$
in their N-body simulations are in good agreement. 

Our semi-analytical model also reproduces evolution curves that
asymptote to the dashed lines.
In addition to the asymptotic values, 
the evolution of the initial buildup stage and
oscillation frequencies in the steady state are also very
similar to Canup \& Ward (2006)'s N-body simulations,
although our results show more regular oscillations.

As mentioned in section 1, our semi-analytical model is
complementary to N-body simulations.
Because our model is faster than N-body simulations 
by many orders of magnitude, we can perform enough number of runs 
for statistical arguments and 
survey different boundary and initial conditions.

\section{RESULTS OF MONTE CARLO SIMULATIONS}

For each set of simulations for Jovian and Saturnian systems (set I and set II;
for details, see Table II), 
we repeat 100 runs with different random numbers for generating the initial locations of proto-satellite seeds and 
values for the parameter $\alpha$ for the disk viscosity and
the timescale $\tau_{\rm dep}$ for exponential decay of mass infall onto the disk.
In addition to the initial locations of the satellite seeds,
we assume log-uniform distributions for 
$\alpha$ in $10^{-2}$--$10^{-3}$
and $\tau_{\rm dep}$ in $(3$--$5)\times 10^6$ years.
The dependence on the value of $\alpha$ is also
very weak as long as $\alpha$ is in this range.
The inner boundary and infall truncation conditions 
are the most important factors that can change the results.  

\subsection{Saturnian Satellite System}

We first show the results of set II, which adopts the same
disk setting as Canup \& Ward (2006)'s N-body simulations.
We apply this setting to simulations of Saturnian satellite systems.
In this case, we found that satellites formed in outer regions 
of the proto-satellite disk repeatedly sweep up inner small satellites,
in a steady state after an initial build up and 
before significant depletion of disk gas.  
Figure~3a 
shows the distributions of the number of final 
surviving satellites with $M > 10^{-5}M_p$ produced 
from 100 simulations for set II.  
Dark gray parts show the runs that produced satellite systems 
in which the largest satellite is mostly icy and $M > 10^{-4}M_p$,
which are analogous to the present Saturnian system.

In about 70\% of the runs, only one body remains 
in the outer regions of the disk. 
Figure~4 shows {\rm the silicate-}dust to gas ratio 
in the disk in the steady state.
If $f_d/f_g = 1$, the ratio is the same as that of 
infalling gas, $1/100$.
This figure shows the disk is highly metal-rich in outer regions.
(We are using the astronomical term ``metal-rich" to denote all condensable solids, including ices, rocks and metals.)
Since in the steady state, $f_g$ is radially constant,
$f_d$ itself is 10--100 times higher in outer regions
than in inner regions, so that
$\tau_{\rm acc}$ is shorter in outer regions (eq.~[13]).
The total mass flux to satellite feeding zones is larger in the outer regions of the disk because the infall mass flux per unit area is assumed to be constant.
From these two facts, large satellites are predominantly formed 
in the outer regions.  
As large satellites migrate from the outer regions, the satellites
accrete satellitesimals in the inner regions during migration
and trap smaller inner satellites into resonances to push them 
to the host planet, because type-I migration is faster for 
heavier satellites (Tanaka et al. 2002).
This process results in
significant depletion of $\Sigma_d$ (equivalently $f_d$) 
in the inner regions.
In the outer regions, the retention of proto-satellites is efficient
against disk gas accretion onto the host planet and the
disk becomes metal-rich ($f_d/f_g \sim 10$-100).

This process repeats until the disk gas is 
so depleted that type-I migration becomes ineffective.  
Since type-I migration is slower in the outer regions and
proto-satellites keep growing near the original locations
until $M$ exceeds $M_{\rm mig}$,
only one large body usually exists in outer regions 
after the disk gas is depleted.  

Several large bodies like the Galilean satellites are found 
under the same simulation setting 
in the previous work (Canup \& Ward 2006).
Our simulations also reproduce the Galilean satellites' analogues
but with a very low probability ($\sim 5$\% for three bodies, 
$\sim 1$\% for four bodies).  
The probability we found may be much lower than the results by
N-body simulations by Canup \& Ward (2006), although
 the evolution of the total satellite mass $M_T$ is very similar (Fig.~2).
As mentioned in section 1,
while the semi-analytical calculations inevitably introduce
approximations, N-body simulations have to employ 
unrealistic initial and/or boundary conditions 
to save computational time.
In the Canup \& Ward (2006)'s N-body simulations, 
infalling materials of dust grains 
are replaced by embryos with the isolation mass,
which is relatively large.
Since evolution of $M_T$ is very similar, our semi-analytical calculations 
and Canup \& Ward (2006)'s N-body simulations are consistent to
each other to some level.
To reconcile the inconsistent details, 
a better calibrated semi-analytical model and
higher-resolution N-body simulations will be needed.

In Fig.~5a, the average mass ($M_s$) and semimajor axis ($a$) 
of the largest satellites 
with their standard deviations obtained from 67 runs that produced 
only one large satellite are plotted and compared with Titan (open circle).
The mass and semimajor axis of the largest bodies agree 
with those of Titan within 1 $\sigma$ standard deviations of 67 runs.  
Furthermore, the largest body contains (1) more than 95\% of 
the total satellite 
system mass in 31 runs out of 67 cases, (2) more than 90\% in 21 runs, 
and (3) less than 90\% in 15 runs.  
The actual significant mass concentration 
in Titan ($\sim 95$\% of the Saturnian satellite system mass) is reproduced
in our simulations.

The colors of the results in Fig.~5
show the average fraction of 
rocky (dark gray) and icy (light gray) components.
Since the final satellites are the last survivors formed 
in the last stage with low disk temperatures due to reduced viscous heating, 
they are generally composed of ice.  
These satellites were usually formed on long timescales 
$\sim$ O$(10^6)$ years, which suggests that Titan is not fully differentiated.
In fact, recent Cassini data indicated that Titan has an inertia factor of 0.34 and is therefore incompletely differentiated just like Callisto (Iess. et al. 2010).

Therefore, our calculations are able to produce many features
(orbits, masses, and compositions) of Saturnian satellite system.
The results of typical successful runs are shown in Fig.~6a.
In these cases, we produced only one large satellite at Titan's location
and some smaller satellites at inner satellites' region.

Castillo-Rogez et al. (2007, 2009) have argued that Saturn's outer regular satellite, Iapetus, must have formed between 3.4 and 5.4 Myrs after the production of the calcium-aluminum inclusions (CAIs) found in meteorites in order to accrete the necessary amount of the short-lived radioisotope $^{26}$Al to ensure its partial differentiation and tidal despinning to its current state of synchronous rotation. 
Mosqueira et al. (2010) also emphasize the important constraint that Iapetus places on any successful model of Saturnian satellite formation.  In future work we plan to address the formation of Iapetus by considering a more detailed evolution model of the outer edge of the proto-satellite disk. 

\subsection{Jovian Satellite System}

For the Jovian case (set I), our numerical simulations produced
4 or 5 large bodies with $M > 10^{-5}M_p$ in about 80\% of the runs
(Fig.~3b).
Dark gray parts show the runs that the produced satellite systems 
in which the inner two bodies are rocky and the outer two are icy satellites.
In the four large satellite cases, the inner three are trapped in mutual 
resonances in all runs, while the outermost one is not trapped 
in a resonance in about half of the runs.
Such resonant trapping of inwardly migrating satellites would reproduce 
the detailed orbital properties of the Galilean satellites (Peale \& Lee 2002).  
Note that in the Saturnian system, this stage may be
followed by the low-$\Sigma_g$ phase without a cavity,
in which all the trapped satellites fall onto the host planet 
and only one dominant icy body formed in the last stage remains,
as explained in section 4.1.  

Figure 3b shows that for set I, 
formation of four large satellites 
is the most likely outcome.
In Fig.~5b, their averaged mass and semimajor axis 
and their standard deviations from 39 runs that produced four large satellites are plotted with Galilean satellites (open circles).
The simulated masses and orbits are consistent with those of
Galilean satellites.

In Fig.~5b, 
the compositions of the simulated satellites are also indicated.
In more than half of the runs, the inner two bodies are 
composed mostly of rocky materials and while the outer two are 
formed mostly of icy materials that have migrated from the regions 
outside the ice line in the proto-satellite disk.  
In the Jovian systems, the final satellites are those formed in 
disks with relatively high infall rates onto the disk,
the disk temperature is high enough 
($T_d \propto F_p^{1/4}$; eq.~[A2]) that 
rocky materials are a major component of satellites formed 
in the inner regions (eq.~[3]).
Outer satellites are formed in the outer regions and migrate inward 
to be trapped in resonances.
Our simulations are consistent with the compositional 
gradient observed in the Galilean satellites.
The results of typical successful runs are shown in
Fig.~6b.
In these cases, we produced four large satellites those
correspond to Galilean satellites for both sizes and locations.
Although some smaller satellites were also formed near Io's location,
they could be merged with Io afterward.
 
The moment of inertia of Callisto suggests that 
its interior is differentiated only partially.
It implies that Callisto must be formed slowly 
such as accretion timescale $5\times 10^5$ years or 
more (Stevenson et al. 1986; Schubert et al. 2004; Barr \& Canup 2008).  
In the case of Jovian system, the outermost large satellite 
corresponding to Callisto is formed on timescales of O($10^5$) years 
in some cases, but generally the timescales would be 
much shorter than the above estimate.

We also performed additional set (set I') of calculations 
with the effects of 
imperfect gap opening, that is,
after the gap opening, we do not truncate the infall but reduce 
the infall rate by a factor of 100.  
In almost all runs, we found that after the gap opening
one additional big body tends to accrete without migration 
in outer regions of the decayed disk on timescales longer 
than O($10^6$) years, which comes from the fact
that $\tau_{\rm acc} \propto f_d^{-1}$.
Because the outermost body hardly migrates, 
it is not captured by a resonance. 
The pronounced difference between Jovian and 
Saturnian satellite systems is preserved, although 
the most probable number of remaining satellites is five.
Thus, if we take into account the effects of imperfect gap opening, 
we found that the formation timescale and orbital 
configuration of the outermost satellite 
in our simulations becomes much more consistent with Callisto.  
To discuss details of 
final orbital configurations and formation timescale 
of the outermost satellite,
we need to take into account the time evolution of radial dependence of 
gas and dust infall rates. 

\subsection{Formation of Saturn's Rings}

We will comment on another pronounced difference between Jovian
and Saturnian systems, planetary rings.
The discovery by the Cassini mission that most of the mass in Saturn's rings is organized into opaque, elongated clumps of ring particles (Colwell et al. 2007, 2009; Hedman et al. 2007) has led to revised estimates of the total mass contained in the rings.  By comparing the amount of light transmitted by the rings during stellar occultations with the transparency of local N-body simulations of the rings, Robbins et al. (2010) estimate that the total mass of Saturn's rings may be twice the mass of the satellite, Mimas, or about 
10$^{-7} M_p$.  The larger mass estimate for Saturn's main rings makes a ring formation scenario that captures a passing large Kuiper belt object less probable and also causes problems for ring formation by the collisional breakup of a former Saturnian satellite unless the impact occurred within the first 750 Myrs after Saturn's formation when the impactor flux at Saturn's orbit was much higher than today (Charnoz et al. 2009).  Spectroscopic observations indicate that Saturn's rings are mainly composed of water ice with less than a few percent by mass of a non-icy component (Cuzzi et al. 2009).  This is a much smaller fraction of rock than is typical for the icy regular satellites in the Saturn system, which presents a problem for ring formation by the breakup of a former satellite unless substantial amounts of silicate rock are hidden under ice regoliths in the dense core of the B ring.  Recent Cassini observations indicate the existence of macroscopic bodies that produce ``propeller"-shaped structures, in the A ring.  The existence of these macroscopic bodies gives new support to a scenario of the ring origin through the catastrophic disruption of a massive progenitor, implying that the ring is old (Tiscareno et al. 2008).
The origin of Saturn's main ring system is therefore still an unsolved question.

Charnoz et al. (2009) showed that the ring may be formed from 
an ancient satellite which was originally in Saturn's 
Roche zone (RZ) and was destroyed by a passing comet.  
(The small satellite itself may have formed outside RZ
and migrated or been scattered into RZ.)
In this model, a key aspect is the survival of the satellite 
against tidal evolution in the planet's RZ.
Since the tidal bulge of the host planet leads to inward migration 
if the satellite's orbital radius ($r$) is below the 
synchronous radius ($R_{\rm sync}$) with the planet's spin.
Typical inward migration timescales are $\sim$ O$(10^8)$ years
for $M \sim 10^{-7} M_p$, which is much shorter than 
the Solar system age, so that for the satellite to survive, 
the condition $R_{\rm sync} < r < R_{\rm RZ}$ is required
(Charnoz et al. 2009), 
where $R_{\rm RZ}$ is the radius of RZ
and is inversely proportional to cube root of
the material density of the satellite ($R_{\rm RZ} \propto \rho_s^{-1/3}$).
For current Jupiter, Saturn, Uranus and Neptune, 
$R_{\rm sync}/R_p = 2.24, 1.86, 3.22$ and 3.36, respectively.
Even if a satellite is an icy body with material density of 
$\rho_s = 1$ g cm$^{-3}$, 
$R_{\rm RZ}$ for Uranus and Neptune are $2.79R_p$ and $2.89 R_p$
that are still smaller than their $R_{\rm sync}$.
Thus, Uranus and Neptune may be unable to maintain 
satellites inside their RZs (Charnoz et al. 2009).

Our results indicate that the inner satellite, which can be 
the seed of the ring, should be 
mixture of ice and rock for Saturnian system, while 
it may be rocky for Jovian system (Fig.~5).  
Since rocky bodies have larger values of $\rho_s$, 
$R_{\rm RZ}$ may be smaller for Jupiter.
For example, 
$R_{\rm RZ} = 1.90 R_p$ with $\rho_s = 1.5$ g cm$^{-3}$ for Saturn, 
which is larger than Saturn's $R_{\rm sync}=1.86R_p$,
while $R_{\rm RZ} = 1.99 R_p$ 
with $\rho_s = 2.5$ g cm$^{-3}$ for Jupiter,
which is smaller than Jupiter's $R_{\rm sync} = 2.24R_p$.
Therefore, Saturn is the most plausible planet to maintain a 
satellite inside the RZ against the tidal evolution.
If the disruption scenario is correct, our result 
{\rm that finally surviving satellites
are formed in a warmer disk for Jovian
system than for Saturnian system}
is consistent with the fact that Saturn 
has a massive ring while Jupiter does not.

\section{CONCLUSION}

We simulated growth and orbital migration of proto-satellites 
in an accreting circum-planetary disk that is modified from
the disk model by Canup \& Ward (2002, 2006), 
in order to address the different architectures between
Jovian and Saturnian satellite systems:
\begin{enumerate}
\item Jovian system has four similar-sized satellites (Galilean
satellites) locked in mean-motion resonances except the outermost
one (Callisto), while Saturnian system has only one big satellite,
Titan, in outer region.
\item Inner two satellites are rocky and outer two are mostly icy
for Jovian system, while Titan is mostly icy.
\end{enumerate}
We modified the semi-analytical model for the population
synthesis model for planet formation developed by (Ida \& Lin 2004, 2008)
to apply the model for satellite formation,
introducing the effect of resonant trapping that characterizes
Galilean satellites. 
The semi-analytical model reproduces the results consistent with
N-body simulations by Canup \& Ward (2006), 
although details such as mass distribution
of surviving satellites are not necessarily consistent
(it is a future work to reconcile the inconsistency).

We considered a coupled system of gas giant planet formation and 
satellite formation around the planet, although planet formation
is not simultaneously simulated.
The fact that multiple gas giant systems are found in extrasolar systems
in addition to our Solar system suggests that gas accretion onto gas giants
is truncated or at least significantly reduced by gap formation
in the proto-planetary disk.
Since Saturn is smaller than Jupiter and
it is theoretically expected that Saturnian core is formed 
later than that of Jovian core, 
we consider a working hypothesis that
{\it Jupiter opened up a gap while Saturn did not.}
Accordingly, we assume the followings: 
\begin{enumerate}
\item Gas infall onto a proto-satellite disk is 
truncated or significantly reduced at the time of gap opening for
Jupiter while that onto Saturnian disk slowly decays
as the proto-planetary disk (the Solar nebula) 
globally decayed on a timescale of $\tau_{\rm dep} \sim 1$-10 Myrs.   
The Jovian disk decays on its viscous diffusion timescale
of $\tau_{\rm diff} \sim 10^3$ years after the gap opening.
\item Based on inference from observations of different
spin periods between CTTSs and WTTSs, 
the proto-satellite disk has an inner cavity in the early, high disk
accretion stage and the cavity vanishes in the subsequent, low disk
accretion stage.
\end{enumerate} 

Since typical type I migration timescales are much
longer than $\tau_{\rm diff}$ and much shorter than $\tau_{\rm dep}$,
the Jovian satellite system may have been ``frozen'' at the time of gap opening
in which disk mass and temperature were relatively high
while the Saturnian satellites may be final survivors in the
last stage in which disk mass and temperature were relatively low.
Thus, we carried out simulations with a cavity and abrupt decay of
disk gas for Jupiter and those without a cavity and with slow
disk depletion on timescales of 1-10 Myrs for Saturn.
Then, we have found the following results:
\begin{enumerate}
\item In Jupiter-condition systems, four or five similar-sized satellites 
are formed in $80\%$ of the runs
and the inner ones tend to be locked in mean-motion resonances 
due to type I migration and its stoppage at the disk inner edge.
In Saturn-condition systems, predominant bodies repeatedly 
accrete in outer regions
and migrate onto the surface of the host planet,
sweeping up inner small satellites. 
In $70\%$ of runs, only one big body remains.
The big bodies have more than $95\%$ of total mass
of remaining bodies in the system
in about half of the runs.
\item Because Jupiter-condition systems are formed in relatively hot disks,
the inner two satellites tend to be rocky and outer ones are ice-rich.
On the other hand, in Saturn-condition systems, the finally surviving bodies form in the longer lasting cold disks and are therefore ice-rich.
Moreover, the surviving massive bodies coresponding to Titan were usually formed on long timescales ($\sim$ Myrs), which is consistent with the incomplete differentiation of Titan's interior.
\end{enumerate}
If we consider a small amount of residual mass infall onto the circum-Jovian disk after the gap opening,
the formation timescale of the outermost satellite is longer
than Myrs, which is consistent with incomplete differentiation of Callisto's interior.
Thus, our simulations produced 
the Galilean satellites' analogues and the Titan's analogues
in each setting with high probability. 

We have demonstrated that
our assumption on gap formation and associated existence of
a disk cavity naturally explain the different
architectures between Jovian and Saturnian satellite systems,
although more detailed fluid dynamical simulations are
needed for the radial dependence and its time evolution of
infall onto proto-satellite disks and gap formation in
proto-planetary disks, and more detailed MHD simulations are
necessary for evolution of a cavity.
The different architectures may be fossil evidence that 
Jupiter opened up a clear gap in the proto-planetary disk 
to terminate its growth.
This result gives deep insights into the mass distribution and multiplicity of extrasolar gas giant planets (Ida \& Lin 2008) 
as well as into the observed bimodal distributions of spin periods of young stars (Herbst \& Mundt 2005).

Notice that the Jovian/Saturnian satellite systems are analogous to extrasolar super-Earth systems, because super-Earths accrete only solid materials and have masses of $\sim (10^{-5}$--$10^{-4})M_{\odot}$.  A high fraction of solar-type stars may have close-in super-Earths (e.g., Mayor et al. 2009), while many systems including the Solar system do not have any close-in super-Earths.  Our results should also give deep insights into the origin of the diversity of super-Earth systems found around other stars.

\acknowledgments

We gratefully acknowledghe the referee's helpful comments.
This work was supported by a JSPS Research Fellowship.  G. R. Stewart was supported by NASA's Outer Planet Research Program and thanks 
the Tokyo Institute of Technology for hosting his visit in 2007.  
S. Ida thanks the University of Colorado at Boulder for hosting his visit 
in 2008.

\appendix

\section{CIRCUM-PLANETARY DISK MODEL}

The growth rate of a proto-satellite depends on the surface density of solid materials of the proto-satellite disk (Ida \& Lin 2004) and its orbital migration rate depends on the surface density of gas (Tanaka et al. 2002).  We adopt the model of an actively-supplied accretion disk by Canup \& Ward (2002, 2006).  The asymptotic disk gas surface density distribution at $r < r_c$, in which the gas from the circum-stellar proto-planetary disk inflows, is given by
\begin{equation}
\Sigma_g = \frac{F_p}{3\pi \nu}\left[1-\frac{4}{5}\sqrt{\frac{r_c}{r_d}}-\frac{1}{5}\left(\frac{r}{r_c}\right)^2\right] \simeq 0.55\frac{F_p}{3\pi\nu},
\end{equation}
where $F_p$ is the total infall rate to whole disk regions at $r < r_c$, $\nu$ is disk gas viscosity, and $r_d$ is the diffused-out disk outer edge, which is $\sim 150R_p$.  The critical radius for the infall, $r_c$, is defined as the orbital radius at which the specific angular momentum of a circular orbit equals that of the inflow average.  
We adopt $r_c = 30R_p$ that is consistent with three-dimensional hydrodynamic simulations with high spatial resolution within the planet's Hill sphere (Machida et al. 2008, Machida 2009).  

The total infall rate is defined with a parameter $\tau_G$ as $F_{p,0} = M_p/\tau_G$ for the steady accretion state.  
According to Canup \& Ward (2006), we adopt $\tau_G = 5\times 10^6$ years for Saturn while $2\times 10^6$ years for Jupiter.  The infall rate decays exponentially with timescale $\tau_{\rm dep} = 3\times 10^6$--$5\times 10^6$ years in the final state of the Saturnian case and abruptly vanishes due to the gap opening in the proto-planetary disk for the Jovian case.  The second and third terms in the bracket of the above equation are correction terms due to outward diffusion.  If they are neglected, $\Sigma_g$ is no other than the distribution of an ordinary steady accretion disk.

The proto-satellite disk is heated by luminosity from the central planet, viscous dissipation, and energy dissipation associated with the difference between the free-fall energy of the incoming gas and that of a Keplerian orbit.  We make the simplification that the viscous dissipation is the dominant energy source, as is typically the case in most of the regular satellite regions (Canup \& Ward 2002).  Then, the photosurface temperature of the proto-satellite disk ($T_d$) is determined by a balance between viscous heating and blackbody radiation from the photosurface,
\begin{equation}
\sigma_{SB}T_d^4 \simeq \frac{9}{8}\Omega_K^2\nu\Sigma_g \simeq \frac{0.55\times 3}{8\pi}\Omega_K^2F_p,
\label{eq:temp_d0}
\end{equation}
where $\Omega_K$ is Keplerian frequency.  
That is, in the steady state with a constant $F_{p,0}$,
\begin{equation}
T_{d,0} \simeq 160\left(\frac{M_p}{M_J}\right)^{1/2}
\left(\frac{\tau_G}{5\times 10^6 \mbox{yrs}}\right)^{-1/4}
\left(\frac{r}{20R_J}\right)^{-3/4}\mbox{K}.
\label{eq:temp_d2}
\end{equation}
After the exponential decay of $F_p$ is imposed,
the disk temperature decreases as
\begin{equation}
T_d = T_{d,0} \exp \left(-\frac{t}{4 \tau_{\rm dep}}\right).
\end{equation}
Note that $T_d$ depends solely on $F_p$ (equivalently, $\tau_G$) 
but is independent of the value of the viscosity $\nu$ (equivalently, $\alpha$) which has 
a large uncertainty. 
After establishment of a steady state, $T_d$ is decreased
from this value according to the decay of $F_p$, 
so proto-satellites are mostly composed of ice in the final state.  

The midplane temperature, $T_c$, is given by $T_c\simeq(1+3\tau/8)^{1/4}T_d$, where $\tau$ is optical depth.  
Because we are concerned with disk temperature
mainly for evolution of the ice line at $T_d \sim 160$ K
that divides compositions of satellitesimals and is located in relatively
outer regions, we assume that $T_c\simeq T_d$ 
to avoid uncertainty in opacity.  

Adopting the alpha prescription for viscosity (Shakura \& Sunyaev 1973), $\nu \simeq \alpha c_s^2/\Omega_K$, the disk gas surface density distribution becomes
\begin{equation}
\Sigma_g \simeq 100f_g\left(\frac{M_p}{M_J}\right)\left(\frac{r}{20R_J}\right)^{-3/4}\mbox{g cm$^{-2}$},
\end{equation}
where
\begin{equation}
f_g \equiv \left(\frac{\alpha}{5\times 10^{-3}}\right)^{-1}\left(\frac{\tau_G}{5\times 10^6\mbox{yrs}}\right)^{-3/4}.
\end{equation}
Since $\nu \propto T_d \propto F_p^{1/4}$,
$\Sigma_g \propto F_p/\nu \propto F_p^{3/4} \propto \tau_G^{-3/4}$.

For simplicity, it is assumed (Canup \& Ward 2002, 2006) that the infall flux per unit area is constant $(F_p/\pi r_c^2)$ at $r < r_c$ and vanishes at $r > r_c$.  The total disk mass inside $r_d$ is  
\begin{equation}
M_{disk} = \int^{r_d}_0\Sigma_g 2\pi rdr \simeq 1\times 10^{-4}f_g\left(\frac{M_p}{M_J}\right)\left(\frac{r_d}{150R_J}\right)^{5/4}M_J.
\end{equation}
We scale the solid surface density with a scaling factor $f_d$ as
\begin{equation}
\Sigma_d = \eta_{\rm ice}f_d\left(\frac{M_p}{M_J}\right)\left(\frac{r}{20R_J}\right)^{-3/4} \mbox{g cm$^{-2}$},
\end{equation}
where $\eta_{\rm ice}$ is an enhancement factor due to condensation of icy grains at $r > r_{\rm ice}$ where disk temperature is lower than 160K; we adopt $\eta_{\rm ice}=3$ for $r>r_{\rm ice}$ and $\eta_{\rm ice}=1$ otherwise. 
Since the satellitesimals' motions are decoupled from gas accretion/diffusion, they stay on-site and $\Sigma_d$ remains zero for $r > r_c$.
The infall rate of solid components per unit area of the disk is given by $(F_p/\pi r_c^2)/(f/\eta_{\rm ice})$, where $f$ is the ratio of gas to rocky dust grains in the solar nebula. 
In our calculations, the radial distribution of solid materials ($\Sigma_d$ or $f_d$) in the proto-satellite disk quickly relaxes to a steady state in which the supply of solid components in the infalling gas is balanced with the removal due to accretion by proto-satellites and their migration.  In the steady state, the rate of increase of $f_d$ due to infall is
\begin{eqnarray}
\frac{df_d}{dt} &=& \frac{d\Sigma_d/dt}{\Sigma_d(f_d=1)} = \frac{(\eta_{\rm ice}/f)(F_p/\pi r_c^2)}{\Sigma_d(f_d=1)} \nonumber \\
 &=& 0.029\left(\frac{M_p}{M_J}\right)^{-2/3}\left(\frac{f}{100}\right)^{-1}\left(\frac{\tau_G}{5\times 10^6\mbox{yrs}}\right)^{-1}\left(\frac{r}{20R_J}\right)^{3/4} \mbox{year$^{-1}$}.
\end{eqnarray}
The increase in the masses of the proto-satellites is subtracted from their feeding zones of width $10r_{\rm H}$.  $f_d$ is assumed to be locally uniform in the feeding zones.  The proto-satellites migrate inward and eventually fall onto the host planets in the presence of disk gas.  Thereby, the total amount of solid materials is decreased by the growth and migration of proto-satellites.  
Our calculation shows that asymptotic values of $f_d$ are 
$\sim 1$ in inner regions and $(10-100)f_g$ in outer regions (Fig.~4), 
which means that the outer disks are generally metal-rich due to the higher retention rate of solid materials compared to the gas.

\section{ACCRETION AND MIGRATION RATES OF PROTO-SATELLITES}

The growth and migration of satellites in circum-planetary proto-satellite disks proceed similarly to those of solid planets in circum-stellar proto-planetary disks that have been studied in detail (Ida \& Lin 2004).  Given a disk model, the accretion and migration rates of proto-satellites are analytically evaluated.  For values of velocity dispersion ($\sigma$) larger than the Hill velocity ($r_{\rm H}\Omega_K$) but smaller than surface escape velocity from satellites, the accretion rate of a proto-satellite with mass $M$ at an orbital radius $r$ is (Ida \& Lin 2004)
\begin{eqnarray}
\dot M &\simeq& C\pi R^2\rho_d\left(\frac{2GM}{R\sigma^2}\right)\sigma \simeq 2C\pi R^2\Sigma_d\Omega_K\left(\frac{GM}{R\sigma^2}\right) \nonumber \\
 &\simeq& 2\pi C\left(\frac{R}{r}\right)\left(\frac{\Sigma_d r^2}{M_p}\right)\left(\frac{r\Omega_K}{\sigma}\right)^2M\Omega_K,
\end{eqnarray}
where $\rho_d$ is the spatial mass density of solid components ($\rho_d \simeq \Sigma_d\Omega_K/\sigma$), $R$ is a physical radius of the satellite, and $C$ is a numerical factor of 2-3 (Stewart \& Ida 2000; Ohtsuki et al. 2002).  In principle, the velocity dispersion $\sigma$ of satellitesimals would be determined by a balance between gas drag damping and stirring by the satellite.  However, because the typical mass of satellitesimals, which determines gas drag damping, accreted by the satellite is not clear, we simply set
\begin{equation}
\frac{\sigma}{r\Omega_K} \simeq \beta\left(\frac{M}{3M_p}\right)^{1/3},
\end{equation}
with $\beta \simeq 10$, following the case of planetesimals in the proto-planetary disks (Ida \& Lin 2004).  The accretion timescale is
\begin{eqnarray}
\tau_{\rm acc} &=& \frac{M}{\dot M} \simeq 0.5\left(\frac{r}{R_p}\right)\left(\frac{\rho}{\rho_p}\right)^{-1/3}\left(\frac{M_p}{\Sigma_d r^2}\right)\left(\frac{M}{M_p}\right)^{1/3}\left(\frac{\beta}{10}\right)^2T_K \nonumber \\
 &\simeq& 10^6 f_d^{-1}\eta_{\rm ice}^{-1}\left(\frac{\rho}{\rho_p}\right)^{-1/3}\left(\frac{M}{10^{-4}M_p}\right)^{1/3}\left(\frac{M_p}{M_J}\right)^{-5/6}\left(\frac{\beta}{10}\right)^2\left(\frac{r}{20R_J}\right)^{5/4} \mbox{years},
\end{eqnarray}
where we used $C=2.5$ and $\rho$ and $\rho_p$ are bulk densities of the satellite and the host planet (we neglect the dependence on $\rho/\rho_p$).  

The type I migration timescale (Tanaka et al. 2002) is
\begin{eqnarray}
\tau_{\rm mig} &=& \frac{r}{\dot r} = \frac{1}{2.7+1.1q_g}\left(\frac{c_s}{r\Omega_K}\right)^2\frac{M_p}{M}\frac{M_p}{r^2\Sigma_g}\Omega_K^{-1} \nonumber \\
 &\simeq& 10^5 \frac{1}{f_g}\left(\frac{M}{10^{-4}M_p}\right)^{-1}\left(\frac{M_p}{M_J}\right)^{-1}\left(\frac{r}{20R_J}\right)^{1/2}\left(\frac{\tau_G}{5\times 10^6\mbox{yrs}}\right)^{-1/4} \mbox{years},
\end{eqnarray}
where $q_g$ is defined by $q_g = -\ln \Sigma_g/\ln r$.

\section{RESONANT TRAPPING}

For two satellites on nearly circular orbits, the expansion of the difference 
in their semimajor axes ($b$) during an encounter is given by 
linear calculations (Goldreich \& Tremaine 1982; Hasegawa \& Nakazawa 1990) as 
$\delta b \simeq 30(b/r_{\rm H})^{-5}r_{\rm H}$.
The Hill radius ($r_{\rm H}$) is defined by $r_{\rm H} = (M/3M_p)^{1/3}r$, where $M$, $M_p$ and $r$ are masses of the proto-satellite and the central planet, and the orbital radius of the proto-satellite, respectively.
Since the scattering occurs at every synodic period 
[$T_{\rm syn} = 2\pi r/((d \Omega/dr)b) \simeq (4\pi r/3 b \Omega_{\rm K}$)], 
\begin{equation} 
\frac{d b}{dt} \simeq \frac{\delta b}{T_{\rm syn}}  
\simeq 7 \left( \frac{b}{r_{\rm H}}\right)^{-4}  
         \left( \frac{r_{\rm H}}{r} \right)^2 v_{\rm K}. 
\label{eq:scat_mig} 
\end{equation} 
The embryos may be trapped in a resonance 
close to the equilibrium value of $b$ that  
satisfies $db/dt = v_{\rm mig}$, 
where $v_{\rm mig}$ is relative radial migration speed 
for the non-interacting case that is given by
equation~(14). 
Then, the equilibrium value is given by 
\begin{equation} 
b_{\rm trap} \simeq 0.29 \left( \frac{M}{10^{-4}M_J} \right)^{1/6}  
         \left( \frac{v_{\rm mig}}{v_{\rm K}} \right)^{-1/4} r_{\rm H}. 
\label{eq:b_trap1} 
\end{equation} 
If the estimated value of $b_{\rm trap}$ is smaller  
than $2\sqrt{3} r_{\rm H}$, the trapping does not actually occur,
because the separation is within their feeding zone 
and $\delta b$ saturates due to the non-linear effect 
(e.g., Ida 1990). 
Substituting eq.~(14) into eq.~(C2),
we find that 
\begin{equation} 
b_{\rm trap} \simeq 16 f_g^{-1/4}
\left( \frac{M}{10^{-4}M_J}\right)^{-1/12}  
\left( \frac{M_p}{M_J}\right)^{-1/4}  
\left( \frac{r}{20R_J}\right)^{1/8} 
\left( \frac{\tau_G}{5 \times 10^6{\rm year}}\right)^{-1/16} 
r_{\rm H}. 
\label{eq:b_trap2} 
\end{equation} 
Since the estimated value of $b_{\rm trap}$ is much greater than
$2\sqrt{3} r_{\rm H}$, we assume perfect trapping
in our simulations.
In our simulations, several satellites are usually trapped, so the final 
trapping location tends to be deeper than the above value
due to collective interactions,
which is also shown in N-body simulations (e.g., Ogihara \& Ida 2009).
We therefore assume that the two approaching proto-satellites are 
trapped at a resonance near $b = 5r_{\rm H}$.

\clearpage

\begin{figure}
\epsscale{1.0}
\plotone{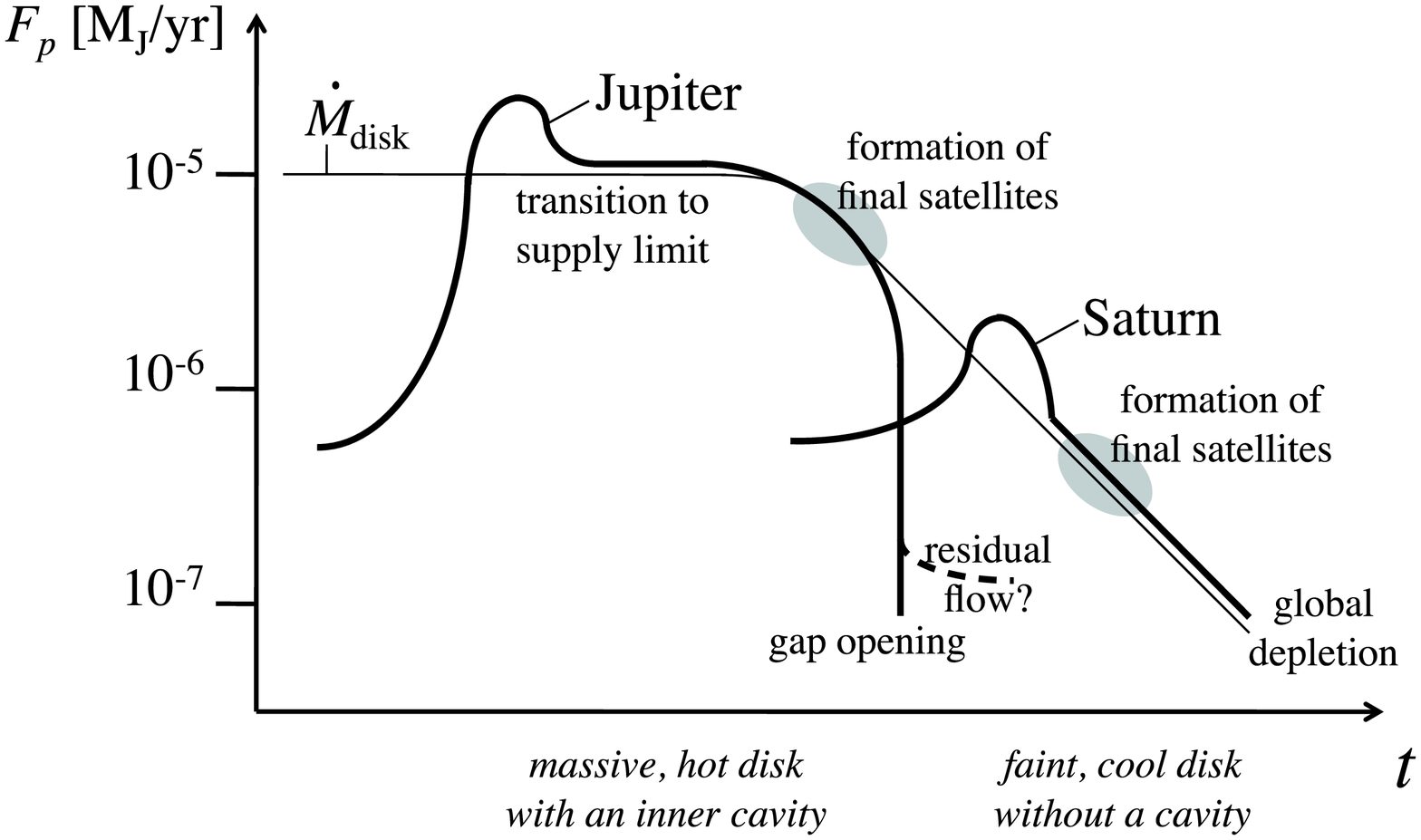}
\caption{Schematic illustration for evolution of 
the infall rate ($F_p$)
onto the proto-satellite disks and timing of satellite
formation in Jovian and Saturnian systems.
Since $\Sigma_g \propto F_p^{3/4}$ and
$T_d \propto F_p^{1/4}$, high $F_p$ means
a massive and hot disk.
}
\label{fig:test}
\end{figure}

\begin{figure}
\epsscale{.80}
\plotone{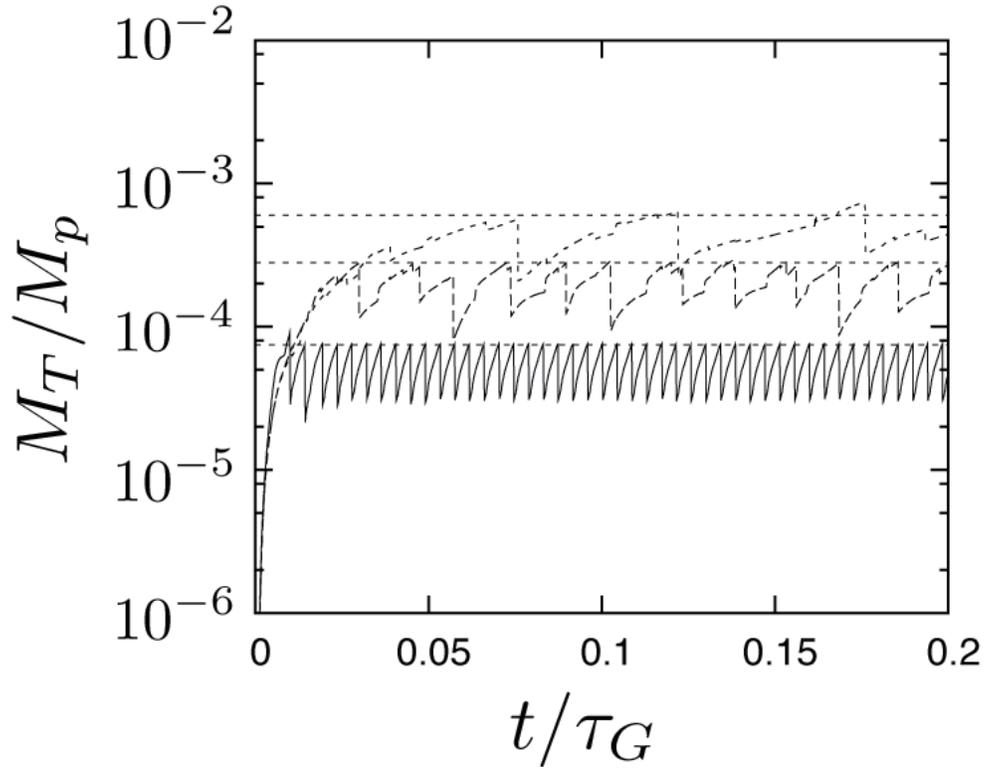}
\caption{Results of calculations with time-constant inflows
to be compared with Canup \& Ward (2006)'s N-body simulation
(Fig.~2 in their paper).
The non-cavity condition is adopted for a Jupiter mass planet.  
The total mass in surviving satellites, $M_T$, 
scaled by the planet's mass ($M_p$) is shown versus time 
scaled to $\tau_G$.  The solid, dashed and dotted lines correspond respectively to calculations with $\alpha = 10^{-4}$, $5\times 10^{-3}$, and $5\times 10^{-2}$.  Dotted horizontal lines are predicted values of $M_T/M_p$ from Eq.~(3) of Canup \& Ward (2006).}
\label{fig:tst2}
\end{figure}

\begin{figure}
\epsscale{1.0}
\plottwo{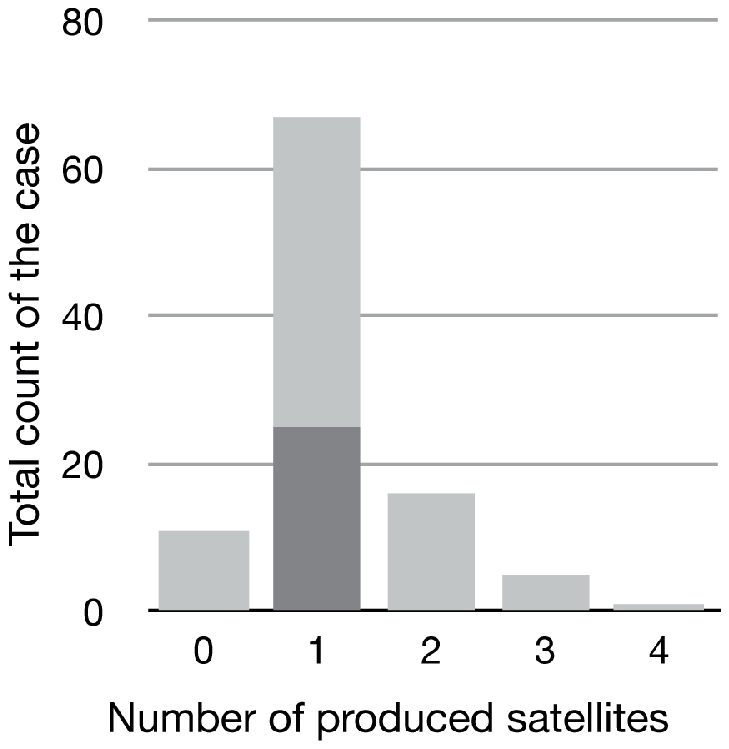}{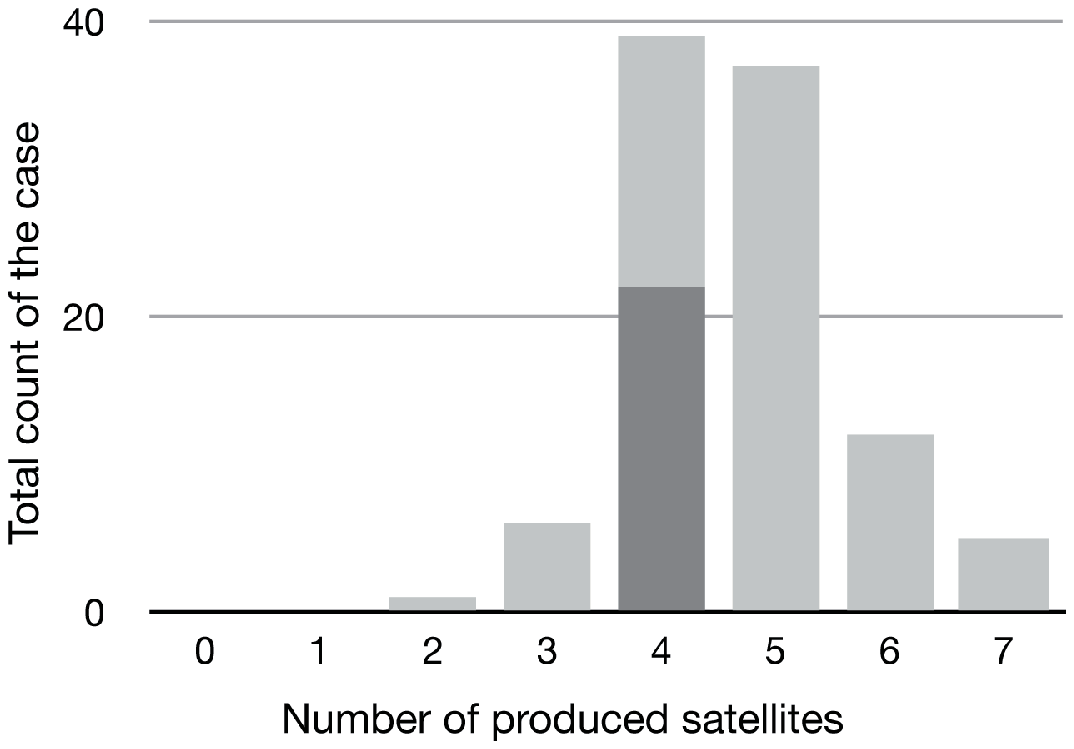}
\caption{Distribution of the number of final surviving satellites 
with $M > 10^{-5}M_p$ produced from 100 simulations for each system, 
(a) Saturnian system and (b) Jovian system.  
Dark gray parts show the runs that the produced satellite systems are analogous to the real one: for the Saturnian system, the largest satellite is icy and $M > 10^{-4}M_p$, while for the Jovian system, the inner two bodies are rocky and the outer two are icy satellites.}
\label{fig:N_satellites}
\end{figure}

\begin{figure}
\epsscale{.80}
\plotone{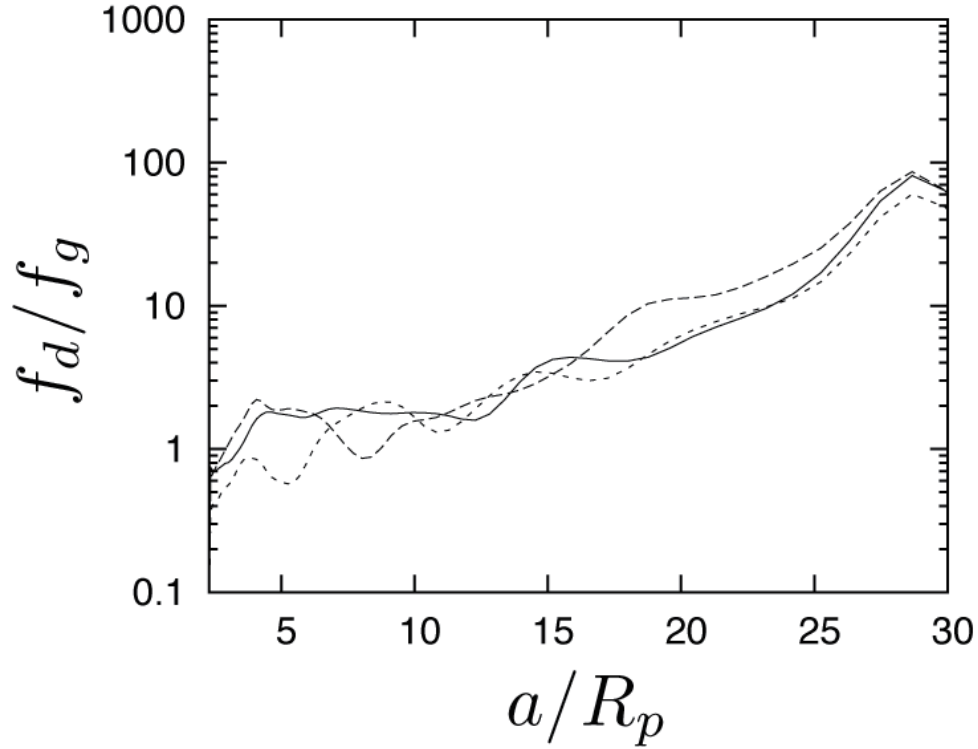}
\caption{Dust to gas ratio in the steady state disk for three typical Saturnian cases.
The solid, dashed and dotted lines correspond respectively to calculations with $N = 15$, 18, and 20.
If $f_d/f_g =1$, the ratio is the same as that of infalling gas, $1/100$. }
\label{fig:f}
\end{figure}

\begin{figure}
\epsscale{1.0}
\plottwo{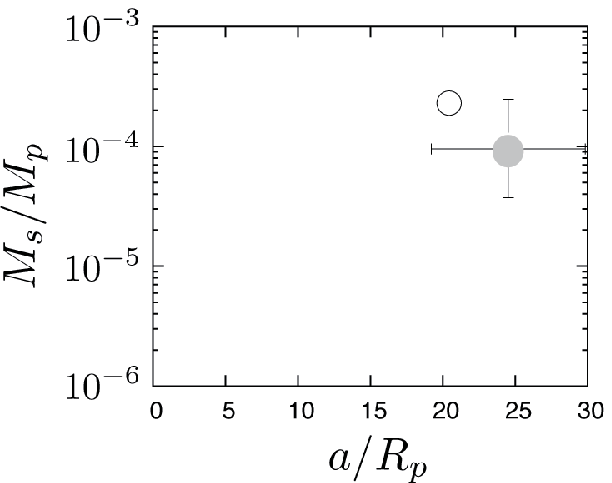}{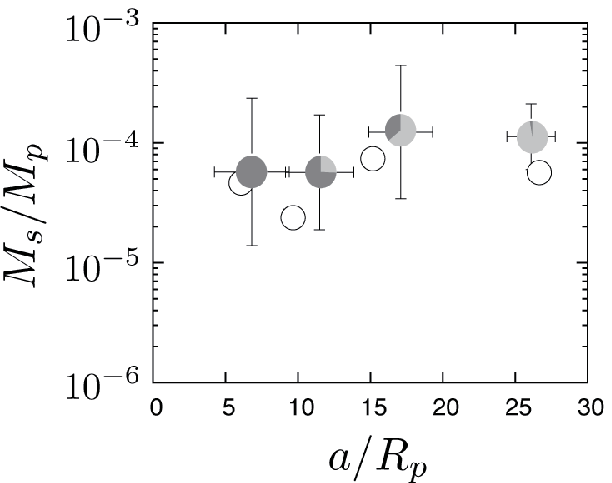}
\caption{Theoretically predicted satellite systems in 
(a) set II corresponding to Saturnian systems
with the data of Titan and (b) set I for Jovian systems with
the data of Galilean satellites.  
For the Saturnian case, 
the average mass ($M_s$) and semimajor axis ($a$) of the 
largest satellites with their standard deviations
are plotted with filled circles with bars for the highest
probability cases (67 runs among 100 runs) 
in which only one large satellite ($M_s > 10^{-5}M_p$) is produced.
The open circle represents Titan.
For the Jovian case, 
the averaged mass-semimajor axis and their standard 
deviations from the highest probability runs (39 runs)
that produced four large satellites are plotted.
Galilean satellites are represented by open circles.
The colors of each plot show the average fraction of rocky (dark gray) and icy (light gray) components of the formed satellites. }
\label{fig:comp}
\end{figure}

\begin{figure}
\epsscale{1.0}
\plottwo{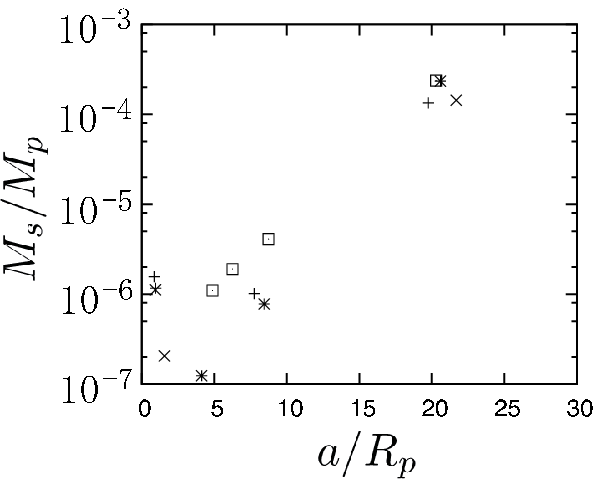}{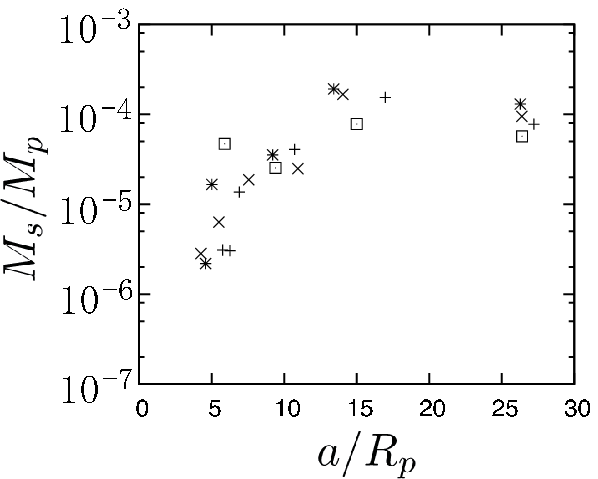}
\caption{Final satellite systems for successful Saturnian and Jovian-like cases in
(a) set II corresponding to Saturnian systems
with the data of large satellites (Tethys, Dione, Dione, and Titan) and (b) set I for Jovian systems with
the data of Galilean satellites.  
The mass ($M_s$) and semimajor axis ($a$) of the produced satellites are plotted
with crosses, pluses, and stars for each successful case, 
while the open squares represent observed satellite systems.
\label{fig5}}
\end{figure}

\clearpage

\begin{table}
\begin{center}
\caption{Satellite properties\tablenotemark{*}. \label{tbl1}}
\begin{tabular}{lcccc}
\\
\tableline\tableline
Satellite & $a/R_p$ & $M/M_p$ [$10^{-5}$]& $\rho$ [g cm$^{-3}$]& $C/MR^2$\\
\tableline
Io & 5.9 & 4.70 & 3.53 & 0.378\\
Europa & 9.4 & 2.53 & 2.99 & 0.346\\
Ganymede & 15.0 & 7.80 & 1.94 & 0.312\\
Callisto & 26.4 & 5.69 & 1.83 & 0.355\\
Titan & 20.3 & 23.7 & 1.88 & 0.34 \\
\tableline
\end{tabular}
\tablenotetext{*}{Source: Schubert et al. 2004); Iess. et al. (2010)}
\end{center}
\end{table}

\clearpage

\begin{table}
\begin{center}
\caption{Simulation settings for Jovian (set I) and Saturnian (set II)
 systems. \label{tbl2}}
\begin{tabular}{l|cccc}
\tableline
\hspace{1em} & $F_{p,0}$ & inner cavity & disk depletion & $\alpha$ \\
\tableline\tableline
Jovian systems (set I) & $M_p/(5\times 10^6$ yrs) & Yes & a & $10^{-3}$--$10^{-2}$ \tablenotemark{*} \\ \tableline
Jovian systems (set I') & $M_p/(5\times 10^6$ yrs) & Yes & b & $10^{-3}$--$10^{-2}$ \\ \tableline
Saturnian systems (set II) & $M_p/(2\times 10^6$ yrs) & No & c & $10^{-3}$--$10^{-2}$ \\ \tableline
\end{tabular}
\tablenotetext{*}{ a log uniform distribution}
\end{center}
a) The evolution of $F_p$ is the same as c) until 
it is truncated at $t = (0.5-0.8)\tau_{\rm dep}$,
which is corresponding to the gap formation. 
\\
b) The same as a) except that $F_{p}$ is
reduced by a factor of 100 instead of complete vanishment
at the gap formation.
\\
c) After establishment of steady state, 
the infall is decayed as $F_p = F_{p,0} \exp(-t/\tau_{\rm dep})$
with $\tau_{\rm dep}=(3$--$5)\times 10^6$ yrs. 
\tablenotemark{*}\\
\end{table}

\end{document}